# Combined stellar structure and atmosphere models for massive stars: Revised ionising fluxes for O stars

Daniel Schaerer

*Geneva Observatory, CH-1290 Sauverny, Switzerland*
*e-mail: schaerer@scsun.unige.ch*

**Abstract.** We present the EUV fluxes derived from the recent "combined stellar structure and atmosphere models" (*CoStar* models). The atmosphere models in particular account for the stellar wind and include non–LTE effects and line blanketing. We provide an extensive set of calculations from tracks between 20 and 120 $M_\odot$ at Z=0.02 (solar metallicity) and Z=0.004. They cover the entire parameter space of O3–B0 stars of all luminosity classes. Model uncertainties are discussed in the light of recent EUV and X-ray observations. We show the importance of using adequate O star ionising fluxes for the interpretation of H II regions, young starbursts and related objects.

## 1. Introduction

In Schaerer et al. (1995ab, hereafter paper I & II) we have presented the first "combined stellar structure and atmosphere models" (hereafter *CoStar*) for massive stars, which consistently treat the entire mass loosing star from the center out to the outer region of the stellar wind. The major advantages of our approach are the following: *1)* it allows to study the effect of a spherically expanding atmosphere and the stellar wind on the interior structure and evolution. *2)* A treatment of the outer boundary conditions allows sound comparisons of stellar parameters for massive stars with strong stellar winds. *3)* The models predict the detailed spectral evolution, including both continuum and line spectra, during the evolutionary phases corresponding to OB, Of/WN, and Wolf–Rayet (WR) stars. *4)* The models represent a first step for future hydrodynamic studies of e.g. WR winds and LBV outbursts, which will require a consistent modeling of both the stellar envelope and the wind.

The *CoStar* models have sofar been applied in detail to the main-sequence (MS) phase (see paper I & II), while in Schaerer (1995) we have studied the effect of the outflowing envelope in WR phases with a simplified method. Our atmosphere models not only treat the non–LTE statistical equilibrium and radiation transfer based on a code from de Koter et al. (1993, 1995). We also include line blanketing with an opacity sampling technique introduced by Schmutz (1991). The calculations thus in particular represent the first models, which account for non–LTE effects, line blanketing, and wind effects and cover the entire parameter space of O and the earliest B stars.

The predicted spectral energy distributions (SEDs) are of special interest for analysis of massive stars, H II regions, young starburst galaxies and related objects. In this contribution we focus on the EUV flux of hot stars, which is strongly affected by non–LTE and wind effects. Our models allow a critical revision of the ionising fluxes of O3–B0 stars.

## 2. Model calculations

The input physics of the *CoStar* models is described in detail in paper I & II and shall therefore not be repeated here. The results from these papers have been extended and they presently cover the MS evolution for stars with initial masses $M_i$ between 20 and 120 $M_\odot$ (Schaerer 1996). This approximately corresponds to spectral types O3 to B0 and all luminosity classes. Calculations are available at solar metallicity (Z=0.02) and Z=0.004 typical for metallicities encountered in H II galaxies. Other metallicities are in preparation. The resulting SEDs are available on request from the author and will be included in the CD-ROM accompanying these proceedings.

### 2.1. Domain of validity and uncertainties

The domain where our atmosphere models are applicable is limited to stars with relatively strong winds due to two assumptions made in the calculations: *1)* The Sobolev approximation made for the line transfer yields good agreement with comoving frame calculations for O and WR stars (de Koter et al. 1993). For weaker winds, however, differences in the level populations will progressively affect the predicted continuum fluxes in particular shortward of the Lyman edge. *2)* Presently our calculations neglect line broadening yielding only a poor treatment of photospheric lines and thus an *underestimate* of blanketing in the photosphere. Therefore our models are actually limited to a domain where wind effects should be dominant for determining the relevant level populations. We estimate the predicted ionising fluxes from the models with $M_i \gtrsim 25\ M_\odot$ to be reliable in this respect. The predictions for the lowest masses are probably more uncertain.

Our calculations do not include the presence of X-rays, which may affect the structure of the atmosphere. However, as the calculations of MacFarlane et al. (1994) show, the importance of this phenomenon depends on the spectral type and increases towards later types. While for O stars X-rays can be regarded as a perturbation on the populations, they play a significant in early B stars (see also Sect 3.1). We thus estimate that the neglect of X-rays should only weakly affect the SED shortward of the He$^+$ edge ($\leq$ 4 Rydberg) for the majority of the models presented in this work. At higher energies, however, the fluxes including X-ray emission may considerably differ (see e.g. Sellmaier et al. 1995).

## 3. Ionising fluxes from wind models and their implications

What is the difference between EUV fluxes from *CoStar* wind models and "usual" plane parallel models ?

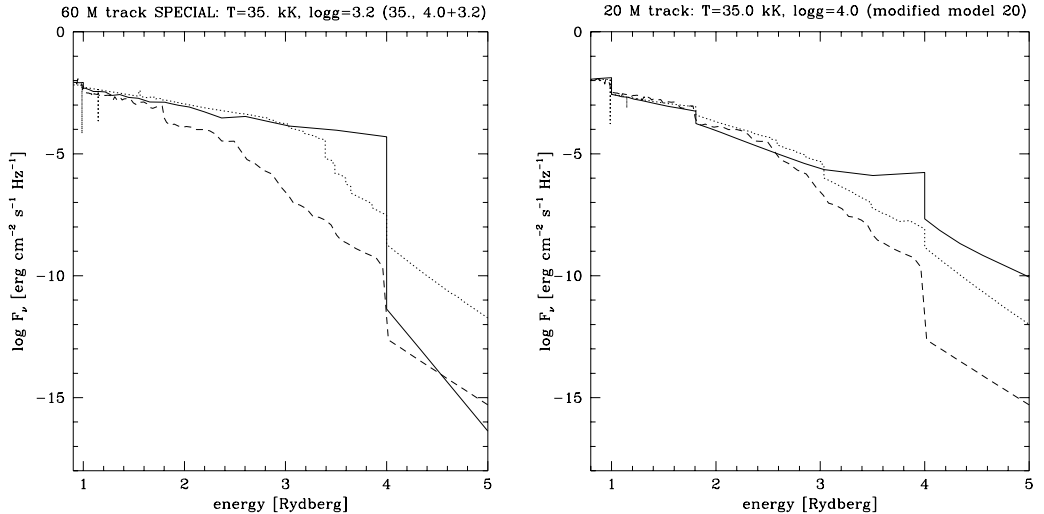

Figure 1. Comparison of emergent EUV fluxes from *CoStar* models (solid), plane parallel non–LTE models of Kunze (1994, dotted) and Kurucz (1991) LTE models (dashed). **a** Supergiant model on a 60 $M_\odot$ track: $(T_{\rm eff}, \log g) = (35\ {\rm kK},\ 3.2)$, **b** Dwarf model on a 20 $M_\odot$ track: $(T_{\rm eff}, \log g) = (35\ {\rm kK},\ 4.0)$

As Gabler et al. (1989) have shown the presence of a stellar wind can lead to a strong depopulation of the ground state of He II, which implies a important increase of the He$^+$ ionising flux (typically 2–3 orders of magnitudes !) over plane parallel models. Thus hot O stars may provide an important contribution to high excitation H II regions (cf. Gabler et al. 1992). Recently it has become clear that similar effects can also work for the He I and even for the H I ground state (Najarro et al. 1995, Paper II). Therefore the He$^\circ$ and the Lyman continuum (although to a lesser extent) can similarly be affected by wind effects.

To illustrate these effects we compare the *CoStar* ionising fluxes with predictions from both LTE and non–LTE line blanketed plane parallel models (Kurucz 1991, Kunze 1994) in Figure 1. The parameters of the models shown roughly correspond to a O9 supergiant and dwarf. The Lyman continuum flux is basically identical for all models. In the He$^\circ$ continuum both non–LTE models (Kunze & *CoStar*) predict larger fluxes for O stars. Typically the number of He$^+$ *ionising photons is increased* by a factor of 1.5 to 2 in *CoStar* models with respect to the Kurucz LTE models. In addition, due to wind effects, the *CoStar* models predict generally a *flatter SED* than plane parallel models in the He$^\circ$ continuum.

Based on *CoStar* models we have revised the ionising photon fluxes for O3–B0 stars using the recent $T_{\rm eff}$–$\log g$ calibration of Vacca et al. (1995). Details are given in Schaerer (1996). The predictions which account for non–LTE, line blanketing, and wind effects, should allow for the first time reliable nebular analysis based on He recombination lines (e.g. He I $\lambda$ 4471, He II $\lambda$ 4686). Nebular studies based on other lines which are sensitive to the detailed SED in the He I and He II continuum will also be modified by the new ionising fluxes. As an example, the exploratory calculations of Sellmaier et al. have shown that
the emergent fluxes from wind models lead to modifications of the ionisation
structure of H II regions and thereby provide a solution to the so-called [Ne III]
problem. To study broader implications for models of H II regions we have
calculated a number of nebular models using the *CoStar* fluxes (see Stasińska
these proceedings and accompanying CD-ROM). The results will be discussed
elsewhere.

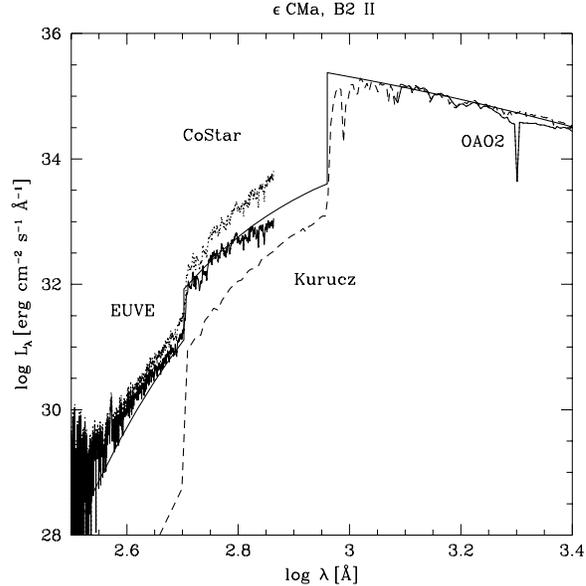

Figure 2.  Far UV and EUV flux of $\epsilon$ CMa. EUVE observations
from Cassinelli et al. (1995a) corrected for an attenuation by $N_{\rm H\,I} =
1.\,10^{18}$ (dotted), and $N_{\rm H\,I} = 5.\,10^{17}$ cm$^{-2}$ (Gry et al. 1995; solid) Model
comparisons: *CoStar* model (solid line, Parameters as Najarro et al.),
Kurucz model (dashed) with $(T_{\rm eff}, \log g) = (21\ {\rm kK}, 3.0)$

### 3.1. Observational tests: comparison with EUVE observations

For obvious reasons direct tests of predictions of atmosphere models shortward
of the Lyman edge are very difficult and rare. Direct observations of O stars –
corresponding to the domain covered by the *CoStar* models – in this wavelength
range do not exist. For B stars, however, the recent EUVE observations of the
B2II giant $\epsilon$ CMa of Cassinelli et al. (1995), provide a unique opportunity in this
respect. In fact these observations have revealed an "EUV excess" of more than
one order of magnitude compared to predictions from both LTE and non–LTE
plane parallel models (Cassinelli et al.). Najarro et al. have shown that wind
effects (as those shown above) can indeed strongly influence the EUV flux even
for the weak wind of $\epsilon$ CMa ($\dot{M} \sim 1.\,10^{-8}\ {\rm M}_\odot {\rm yr}^{-1}$).

In Figure 2 we show preliminary results obtained for $\epsilon$ CMa (cf. Schaerer
1996). The agreement with the observations (especially if corrected for the low
neutral gas column density derived recently by Gry et al. 1995) is surprisingly good and, in fact, somewhat fortuitous ! It must indeed be noted that due to the approximations made in our present models (see Sect. 2.1) the calculations cannot be extrapolated to B stars without caution. In particular the results strongly depend on the treatment of line blanketing in the photosphere and the temperature structure in the photosphere-wind transition zone. It appears that early B stars are probably the most challenging hot stars to model, since they lie in the domain where stellar winds become weak, photospheric processes (including blanketing) gain of importance, and the impact of X-rays becomes determinant. Models which accurately treat all these phenomena seem thus to be required for a better representation of B stars. In this picture one would, however, expect the current models to be sufficiently *reliable for O stars* and types later than B, which are outside of the "transition domain" of mid B stars.

**Acknowledgments.** I express my gratitude to Alex de Koter, Werner Schmutz and André Maeder who have greatly contributed to important parts of the current *CoStar* models. Dietmar Kunze kindly provided results from his atmosphere calculations. This work was supported by the Swiss National Foundation of Scientific Research.

# References

Cassinelli J.P. et al. 1995, ApJ 438, 932

Gabler R., Gabler A., Kudritzki R.P., Puls J., Pauldrach A. 1989, A&A 226, 162

Gabler R., Gabler A., Kudritzki R.P., Mendez R.H. 1992, A&A 265, 656

Gry C., Lemonon L., Vidal-Madjar A., Lemoine M., Ferlet R. 1995, A&A 302, 497

de Koter A., Lamers H.J.G.L.M., Schmutz W. 1995, A&A , in press

de Koter A., Schmutz W., Lamers H.J.G.L.M. 1993, A&A 277, 561

Kunze, D. 1994, PhD thesis, Ludwig-Maximilians-Universität, Munich, Germany

Kurucz R.L. 1991, in "Stellar Atmospheres: Beyond Classical Models", NATO ASI Series C, Vol. 341, Eds. L.Crivellari, I.Hubeny, D.G.Hummer, p. 441

MacFarlane J.J., Cohen D.H., Wang P. 1994, ApJ 437, 351

Najarro F., Kudritzki R.P., J.P. Cassinelli. O. Stahl, Hillier D.J. 1995, A&A in press

Schaerer D. 1995, A&A, in press

Schaerer D. 1996, A&A, in preparation

Schaerer D., de Koter A., Schmutz W., Maeder A. 1995ab, A&A, in press (Paper I & II)

Schmutz W. 1991, in "Stellar Atmospheres: Beyond Classical Models", Eds. Crivellari, L., Hubeny, I., Hummer, D.G., NATO ASI Series C, Vol. 341, p. 191

Sellmaier F., Yamamoto T., Pauldrach A.W.A., Rubin H. 1995, A&A , submitted

Vacca W.D., Garmany C.D., Shull J.M. 1995, ApJ , in press